# Oxidation Resistant Germanium Nanowires: Bulk Synthesis, Long Chain Alkanethiol Functionalization and Langmuir-Blodgett Assembly

Dunwei Wang[†], Ying-Lan Chang[‡], Zhuang Liu[†], and Hongjie Dai[†*]

†.   Department of Chemistry, Stanford University, Stanford, CA 94305

‡.   Agilent Laboratories, Agilent Technologies, Inc. 3500 Deer Creek Road, Palo Alto, CA 94304.

*Corresponding author: hdai@stanford.edu

## Abstract

A simple method is developed to synthesize gram quantities of uniform Ge nanowires (GeNWs) by chemical vapor deposition on preformed, monodispersed seed-particles loaded onto high surface area silica support. Various chemical functionalization schemes are investigated to passivate the GeNW surfaces using alkanethiols and alkyl Grignard reactions. The stability of functionalization against oxidation of germanium for various alkyl chain lengths is elucidated by X-ray photoelectron spectroscopy.  Among all schemes tested, long chain alkanethiols ($\geq C_{12}$) are found to impart the most stable GeNW passivation against oxidation upon extended exposure to ambient air. Further, the chemically functionalized oxidation-resistant nanowires are soluble in organic solvents and can be readily assembled into close-packed Langmuir-Blodgett films potentially useful for future high performance electronic devices.

**Introduction**

Germanium is an electronic material with renewed interest for future computing owing to high carrier mobilities than silicon.[1-8] However, several major challenges should be met in order for Ge to become a widely useful material for high performance electronics. The cost to prepare single crystal Ge wafers is high since Ge is a naturally scarce element and mass production is lacking. Methods for efficient formation and usage of crystal Ge should be developed to circumvent this problem. A more fundamental issue is that Ge easily forms unstable oxides on the surfaces.[9] $GeO_2$ is soluble in $H_2O$ and the Ge/$GeO_2$ interface has considerably high surface states.[6, 10] Applications of Ge will be limited unless $GeO_2$-free surfaces and interfaces can be obtained by developing robust chemical passivation of Ge to prevent oxidation.

Among various approaches to high quality single crystalline Ge is simple and low thermal budget synthesis of Ge nanowires at temperatures down to 300°C using Au nanoparticles as seeds.[4,7] For such chemical bottom-up approach however, it is necessary to be able to synthesize bulk quantities of GeNWs and assemble them from random entangled forms into ordered structures to facilitate integration into electronic devices. In terms of surface passivation, various functionalization chemistry for oxidation prevention has been investigated in the past for planar Ge surfaces with well defined crystallographic orientations.[11-15] Relative recent is the passivation effort on GeNWs.[15] It remains to be seen whether the various passivation chemistry for planar surfaces can be transferred to curved GeNW surfaces. Also, the efficacy of GeNW passivation by different chemical schemes for molecules with different chain lengths has not been systematically explored.

Obtaining GeNWs with truly robust stability against oxidation (e.g., during extended exposure to *ambient* air) remains a challenge.

In this article, we present an effort to address these issues and meet the various challenges posed to the synthesis, passivation and assembly of GeNWs. First, we demonstrate chemical vapor deposition (CVD) synthesis of GeNWs at large scales using supported Au nanoparticle seeds on high surface area silica. Secondly, a robust surface passivation scheme is identified to stabilize the surfaces of GeNWs using self-assembled crystalline monolayers of long chain alkanethiols (≥12 carbon) for the first time. This imparts chemical stability to GeNWs against oxidation upon exposure to ambient air for ~1 day. The surface functionalized and passivated GeNWs exhibit high solubility in organic solvents and can be readily employed to form Langmuir-Blodgett films on various substrates. The current work thus achieves chemically stable (for ~ 1 day in ambient air) GeNWs in organized forms at large scales, which should facilitate the bottom-up approach to high performance electronics based on Ge.

**Results and discussions**

**Bulk synthesis of GeNWs and purification.** Recently we reported understanding and optimization of GeNW synthesis to achieve nearly 100% yield with every Au nanoparticle capable of nucleating and growing a nanowire.[7] The optimized growth condition was utilized here to produce large quantities of GeNWs on high surface area support materials (Fig. 1, see supporting materials for details). A typical SEM image of as-grown GeNWs from silica supported Au seeds is shown in Fig. 2a. Large numbers of GeNWs emanating from silica particles are observed. After purification by detaching GeNWs from silica support via sonication and removal of silica by centrifugation and HF etching, we obtain large quantities of pure GeNWs free of any silica residue (Fig. 2b). Using monodispersed ~ 2mg of 10±2nm and 20±3nm Au seeds loaded onto silica, we produced ~ 0.8g 13.5±2nm and 0.4g 23.9±3.7nm GeNWs, respectively.

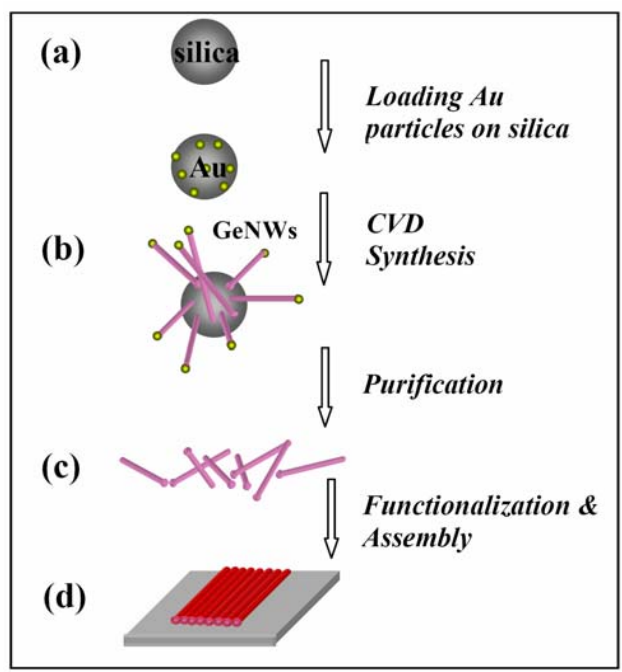

**Figure 1.** Schematic of the approach used for bulk synthesis, purification, functionalization and assembly of GeNWs.

Our bulk synthesis of GeNWs is simple and the idea of loading pre-formed Au nanoparticles onto APTES modified high surface-area silica is novel.  By so doing, we are able to obtain large numbers of *monodispersed* growth seeds for the synthesis of bulk quantities of NWs that are uniform in diameter. Conventional methods employ metal salt precursors impregnated onto silica support and rely on heat treatment to form supported seed catalyst particles. We found that such methods were unable to produce monodispersed Au seeds due to uncontrolled metal aggregation during annealing. Our method of loading preformed monodispersed Au particles onto support materials for NW synthesis can also be generalized for bulk CVD synthesis of other types of nanowires with high uniformity. Previously, bulk quantities of semiconductor nanowires have been reported by using laser ablation[16, 17] or high pressure and temperature supercritical fluid solution-phase reactions.[3, 18] Compared to these methods, our bulk CVD synthesis has advantages in simplicity, low thermal budget and high uniformity of nanowires.

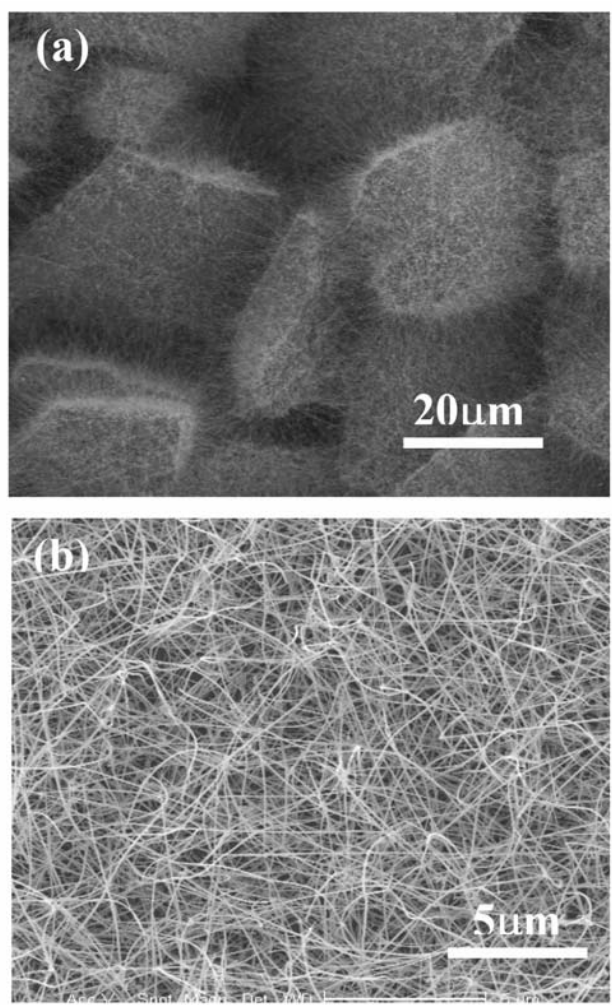

**Figure 2.** Bulk synthesis and purification of Ge nanowires. (a) A SEM image of high yield of GeNWs grown from Au nanoparticles seeds loaded onto silica particles. (b) A SEM image of purified GeNWs after removal of the silica support.

**Passivation of GeNWs via alkanethiols and Grignard reactions**. A major challenge to Ge is the easily oxidized surface[9] and large numbers of Ge/GeO$_2$ interface states rendering poor electrical properties to Ge devices.[10] We have found recently significant band bending due to Fermi level pinning by surface states of oxidized GeNWs.[6] Various schemes of functionalization and passivation have been reported previously[11-14, 19] to impart oxidation resistance to planar Ge surfaces. In the current work, we focused on investigating GeNW functionalization by various chain-length

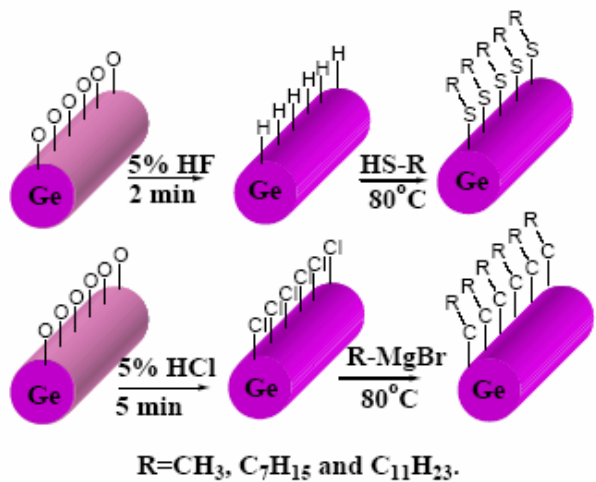

R=CH$_3$, C$_7$H$_{15}$ and C$_{11}$H$_{23}$.

**Figure 3.** GeNW functionalization schemes using alkanethiols (top) and Grignard reactions (bottom).

alkanethiols (similar to a previous reaction[15] but extending to longer chain alkanethiols) and by Grignard reactions,[12] as depicted in Fig. 3. For both the thiol and Grignard reactions, GeNWs were first treated with diluted HF or HCl to remove native oxides and terminate the surfaces with H (for alkanethiol reactions) or Cl (for Grignard reactions). XPS revealed that the as-cleaned GeNWs after these treatments are indeed oxide free (Fig. 4a and 4b, 'as-cleaned' curves), suggesting that Ge NWs terminated with H and Cl are stable for at least 10 min in ambient air during sample transfer into the vacuum chamber for XPS measurements. Controlled exposure of functionalized GeNW to ambient air was carried out to investigate the oxidization behavior of GeNWs by high resolution XPS. Both Ge 2p and 3d binding energies were recorded to monitor the degree of oxidation. Owing to the higher binding energy and smaller escape depth of Ge 2p electrons (1.07nm) compared with 3d electrons (2.11nm), a more sensitive probe to the Ge surface properties can be achieved with Ge 2p spectrum.[20] Without any functionalization, H- and Cl-terminated GeNWs are unstable upon exposure to ambient air for 2h, evidenced by the appearance of a shoulder near the $Ge^0$ 2p peak due to oxidation. After 24 h air exposure, the shoulder grows into an obvious peak corresponding to $Ge^{4+}$ accompanied by a blue shift of the $Ge^0$ peak (see Fig. 4a,4b 'control' curves) due to band bending caused by Fermi level pinning by surface states.[6] In stark contrast, GeNWs functionalized by both alkanethiols and the Grignard reactions (for $C_2$, $C_8$ and $C_{12}$) exhibit much reduced oxide signals than the unfunctionalized ones after 1 day air exposure (Fig. 4a and 4b, '24 hour' curves), pointing to the expected passivation effect. The most stable passivation of GeNWs is achieved with $C_{12}$ alkanethiol (Fig. 4b top curves). The C$_{12}$-thiol coated GeNWs exhibit negligible Ge-oxide signal after 1 day air exposure.

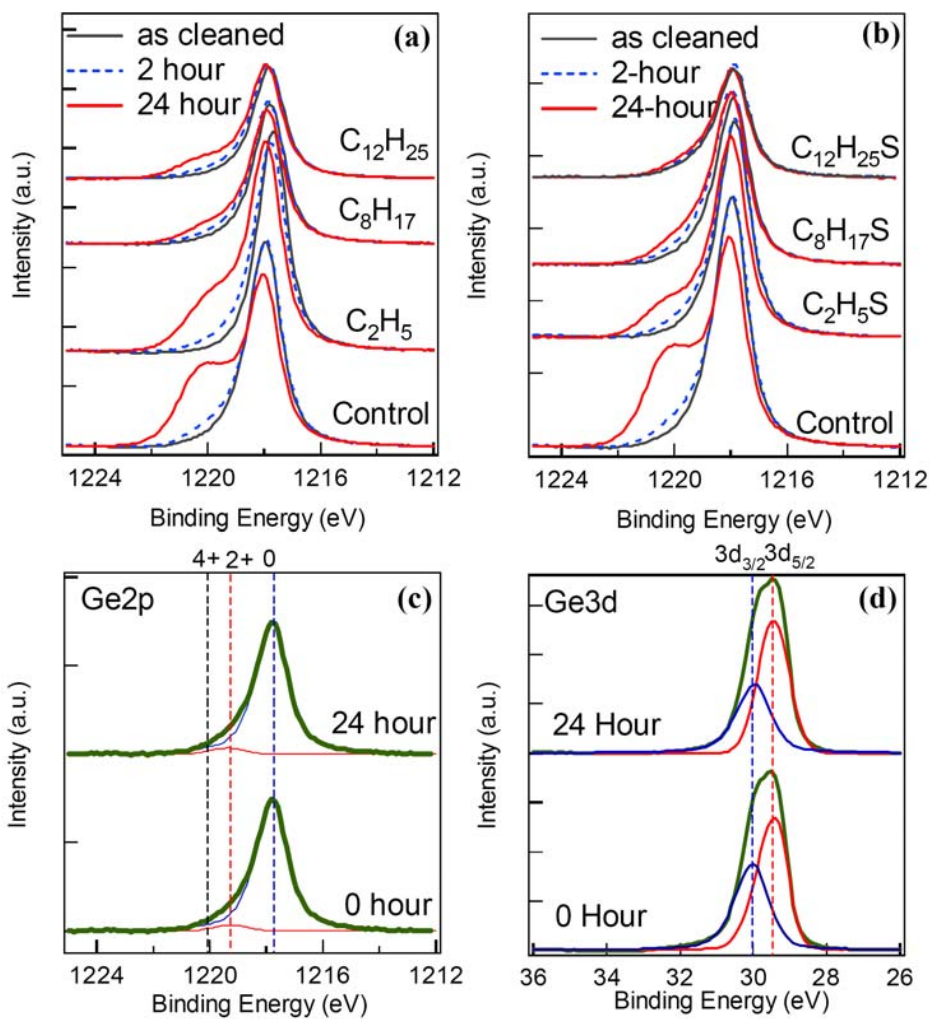

**Figure 4.** XPS data of GeNWs. (a) Ge 2p spectra for GeNWs functionalized by C$_2$, C$_8$ and C$_{12}$ alkyl chains via the Grignard reaction. (b) Ge 2p spectra for GeNWs functionalized by C$_2$, C$_8$ and C$_{12}$ alkanethiols. (c) Peak fitting for Ge 2p spectra before and after 24 h air exposure. A small peak due to Ge$^{2+}$ is discerned by data fitting with intensity remaining unchanged after 24 h exposure to ambient air. (d) Peak fitting for Ge 3d spectra before and after 24 h air exposure. No GeO peak is discerned here due to lower sensitivity of 3d spectra to surface properties than 2p spectra.

XPS data obtained with Ge 2p electrons can unveil small changes not discernable in 3d data. For GeNWs right after C$_{12}$H$_{25}$SH functionalization, peak fitting of the Ge3d spectrum identified no Ge-oxide signal (Fig. 4d). However, similar analysis performed on the Ge 2p spectrum clearly identified a small peak corresponding to $Ge^{2+}$ (from GeO) (Fig. 4c). This suggested minor oxidation of the as-cleaned GeNWs during sample transfer between acid cleaning and functionalization. The intensity of the peak exhibited no change after 1 day exposure in ambient air (Fig. 4c) owing to the $C_{12}H_{25}SH$ passivation layer. The result here strongly suggests that Ge 2p spectra should be used to analyze surface functionalization effects by various molecules to afford high sensitivity.

**Chain length effect.** We carried out a systematic investigation of the alkyl chain length ($C_2$-$C_{12}$) effect to the oxidation resistance of GeNWs after functionalization by thiols and Grignard reactions. The degree of oxidation (to $GeO_2$) of GeNWs was extracted from peak fitting of Ge 2p binding energies and plotted against the exposure time to ambient air in Fig. 5. For alkanethiol functionalized GeNWs, the stability of the wires monotonically increases with the alkyl chain length, with the dodecanethiol ($C_{12}H_{25}SH$) functionalized GeNWs exhibiting essentially oxide free surfaces after 1 day air exposure (Fig. 5a). Similar dependence of oxidation resistance to alkyl chain length has been reported previously for metal surfaces (e.g., Cu) with self-assembled monolayers (SAM). As for the SAM protected metals, we suggest that the enhanced oxidation resistance of GeNWs by longer chain alkanethiols is owing to the better packing of the molecules resulted from stronger inter-chain van der Waals interactions and thus more crystalline and denser molecular films formed on the GeNW surfaces. This affords better blocking of oxygen and therefore higher oxidation resistance.[23] Since the GeNWs employed in the current work are ~ 10-20 nm in diameter and significantly larger than the molecular chain length (up to ~1nm for $C_{12}$), dense crystalline alkanethiol SAMs can still be formed on these relatively large wires. For NWs with diameters approaching 1-2 nm, different passivation strategies will be necessary since close packing of ~1 nm long molecules on the NWs will be difficult due to the high curvature of the wires.

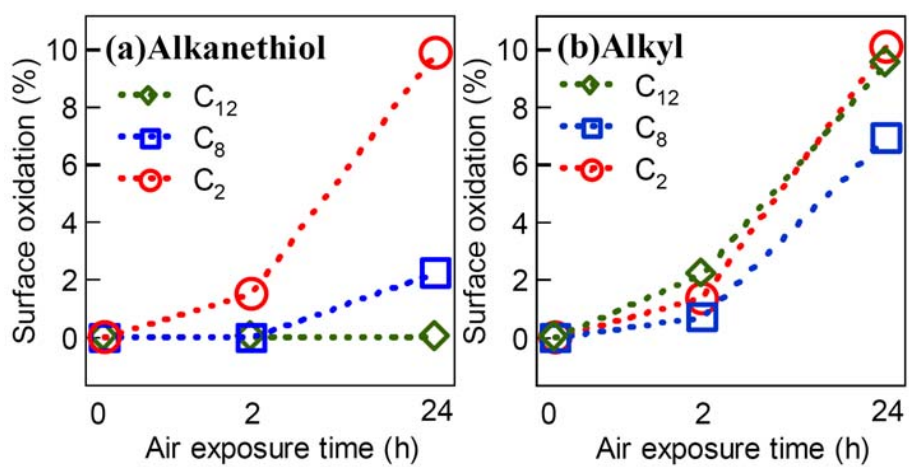

**Figure 5.** Oxidation behavior of various functionalized GeNWs. The percentage of surface oxides ($GeO_2$) is extracted from peak fitting of Ge 2p spectra. (a) Data for various alkanethiols functionalized NWs. (b) Data for various alkyl Grignard reactions. The $C_{12}$-thiol functionalization in (a) gives the most effective passivation .

The overall $Ge^0$ signal intensity monotonically decreases for GeNWs modified by longer chain alkyl groups via both thiol and Grignard reactions (under the same data collection condition, Fig. 4a and 4b). This is consistent with the formation of thicker molecular layers for longer alkyl chains. We also carried out TEM imaging of the molecular passivated GeNWs and found that the thickness of the coating on the nanowires consistently increased with the alkyl chain length (Fig. 6). For GeNWs reacted with $C_2H_5SH$, $C_8H_{17}SH$ and $C_{12}H_{25}SH$, the thickness of the coating layers over the GeNW crystalline lattices were ~0 Å, ~7Å and ~11Å respectively.

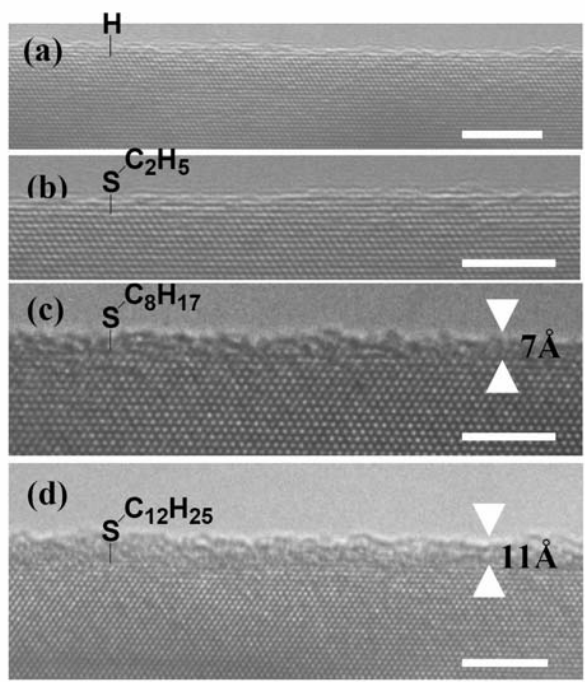

**Figure 6.** TEM images of GeNWs functionalized with thiols of different alkyl chain lengths. (a) As-cleaned GeNWs without functionalization. (b) Ethanethiol ($C_2$), (c) octanethiol ($C_8$) and (d) dodecanethiol ($C_{12}$) thiol functionalized wires. Scale bars in all pictures: 5nm.

The chain length dependence of GeNW stability after alkyl Grignard functionalization somewhat differs from those functionalized by alkanethiols. While the $C_8$- and $C_{12}$- Grignard functionalized GeNWs are significantly more oxidation resistant than the $C_2$ Grignard functionalized ones, the $C_8$ and $C_{12}$ passivations are rather similar (Fig.4a and Fig. 5). We speculate that this could be a result of interplay between the molecular packing/film thickness effect and reactivity differences of molecules with various chain lengths. Longer chain molecules may exhibit lower reactivity with the GeNWs towards Ge-C bond formation, giving little advantage for $C_{12}$- passivation over $C_8$ passivation.

**Relative stability of alkanethiol and alkyl Grignard functionalized GeNWs.**
Unlike the $C_{12}$-thiol passivation, none of the $C_2$, $C_8$ and $C_{12}$-Grignard functionalization investigated here imparted high oxidation resistance to GeNWs (Fig. 4 and Fig. 5). This is in agreement with the result of Korgel et al.[15] but seemingly differs from previous reports that alkyl monolayers on planar Ge (111) and (100) surfaces via Ge-C bonding provide better passivation than alkanethiols.[12, 14, 19] It has been suggested that such stability stems from stronger Ge-C bonding than Ge-S bonding.[19] Nevertheless, in the literature, systematic experimental and theoretical data of Ge-C vs. Ge-S bonding properties for various well-defined Ge crystal surfaces appears lacking. In one reference, we find that that the bonding energy of Ge-C (460kJ/mol) is suggested to be lower than that of Ge-S (534kJ/mol),[24] opposite to the suggestions based on functionalization stability studies. Therefore, the relative bonding strengths for Ge-C and Ge-S remains to be fully elucidated and may not apply to interpreting the relative stability of thiol and alkyl functionalized GeNWs.

Another likely factor is the different reactivity of molecules employed in our functionalization and passivation. It is known that long chain Grignard reactions with Ge require more rigorous reaction conditions or longer times to complete than alkanethiols[12] due to the relatively low reactivity of long chain Grignard reagents. It is therefore probable that our $C_8$ and $C_{12}$ Grignard reactions afford lower coverage on GeNWs than the same length alkanethiols and thus giving less passivation effect.

**Langmuir-Blodgett film.** The Langmuir-Blodgett (LB) technique has been widely applied for molecular films and for various nanomaterials such as nanoparticles, nanorods and nanowires.[25-30] The GeNWs functionalized by both alkanethiols and alkyls form uniform and stable suspensions in organic solvents such as chlorobenzene and chloroform (Fig. 7a inset). This is one of the key elements for successful LB film assembly of GeNWs. Without functionalization, GeNWs do not form stable suspensions in these organic solvents and sink into water due to the hydrophilic oxide surfaces. Upon adding functionalized GeNW suspensions drop-wise to a subphase of ethanol/water in a LB trough (see supporting materials for experimental details), we observed that the droplets spread out quickly on the water surface.  The organic solvent quickly vaporized to leave the functionalized GeNWs floating on the water surfaces. These floating GeNWs formed a close-packed dense film upon compressing with the nanowires oriented perpendicular to compressing direction or parallel to the edge of the trough (Fig.7), and were transferred onto various substrates and ready for characterization or integration into device structures. The subphase of ethanol in $H_2O$ was another key to our successful GeNW LB film formation. We found that without the addition of 5% ethanol in water, monolayer formation of GeNWs on a pure $H_2O$ surface was difficult as the nanowires tended to overlap and form bundles due to strong hydrophobic interactions between functionalized GeNWs. As a result, multilayers of GeNWs or overlapping wires without good alignment were obtained after transferring to solid substrates.  On the other hand, high concentrations of ethanol in the subphase (>10%) is also undesirable since the functionalized nanowires tended to submerge into the solution without floating on surface.

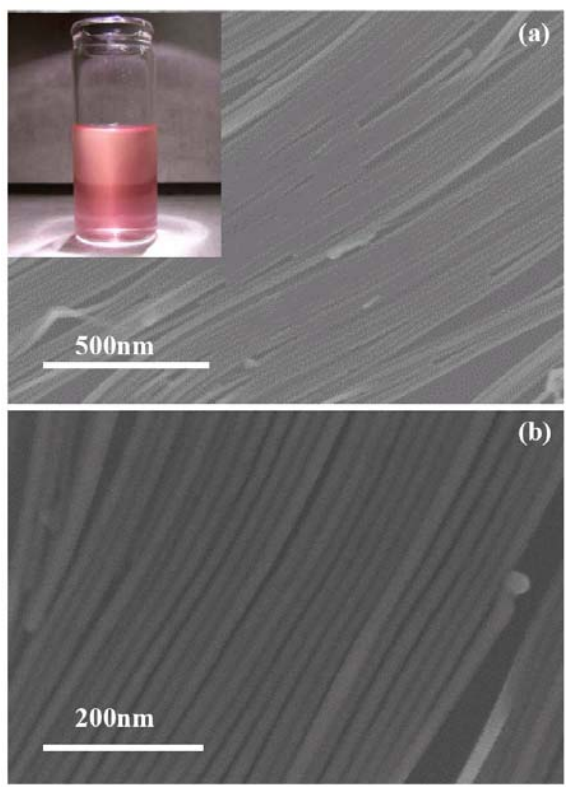

**Figure 7.** Langmuir-Blodgett film of GeNWs. (a) SEM image of a GeNW film with dodecanethiol ($C_{12}$) functionalization. Inset: photo of a GeNW suspension in chloroform. (b) SEM image of a GeNW film with octyl ($C_8$) functionalization.

Simple calculation suggests that ~ 0.7mg GeNWs (20nm in diameter) is needed to form a LB film to cover a full 4-inch wafer. This quantity can be easily afforded by our bulk synthesis that yields ~0.4g of pure GeNWs in a single growth run. Thus, the approach presented here (Fig.1) is promising for large scale productions of LB films of high quality single crystal GeNWs with close packing and excellent alignment. These NW films can be used as wafer materials for fabrication of various types of devices such as field effect transistors with channel length parallel to the nanowire orientations.

**Conclusions**

We have demonstrated the synthesis and purification of bulk quantities of high quality single crystalline GeNWs by CVD on high surface-area silica supported monodispersed Au seeds. We investigated various surface passivation schemes for GeNWs and found that the best method among all tested schemes to protect GeNWs from oxidation in ambient air is to use self-assembled crystalline monolayers of long chain alkanethiols ($\geq$12 carbon). By no means can this passivation method provide indefinite protection of Ge nanowires against oxidation especially under harsher conditions and the quest for optimum Ge passivation will continue. Our surface functionalized GeNWs exhibit high solubility in organic solvents and can be readily employed to form Langmuir-Blodgett films on various substrates. Thus, we achieve oxidation resistant GeNWs in organized forms at large scales, which should significantly facilitate the bottom-up approach to high performance electronics based on single crystal germanium.

**Acknowledgements.**

We would like to thank Profs. Nathan Lewis, Chris Chidsey and Nicholas Melosh for insightful discussions and suggestions. This work was supported by Stanford INMP, a DARPA 3D Program and SRC/AMD.

TOC entry:

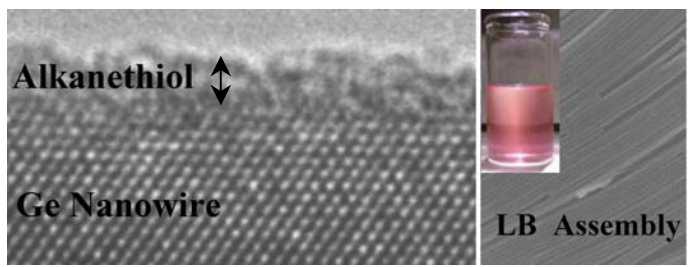

**Oxidation Resistant Germanium Nanowires: Bulk Synthesis, Long Chain**

**Alkanethiol Functionalization and Langmuir-Blodgett Assembly**

Dunwei Wang[†], Ying-Lan Chang[‡], Zhuang Liu[†], and Hongjie Dai[†*]

†.   Department of Chemistry, Stanford University, Stanford, CA 94305

‡.   Agilent Laboratories, Agilent Technologies, Inc. 3500 Deer Creek Road, Palo Alto, CA 94304.

*Corresponding author: hdai@stanford.edu

<u>**Supporting Materials.**</u>

**Materials and Methods**

**Materials.** High surface area ($100 \text{m}^2$/gram) silica particles purchased from Sorbent Technologies, Inc. (Atlanta, GA) were used as support for Au seed nanoparticles (Ted Pella, Redding, CA) for GeNW growth. Functionalization reagents of dodecylmagnesium bromide ($C_{12}H_{25}MgBr$, 1.0M in diethyl ether), octylmagnesium bromide ($C_8H_{17}MgBr$, 2.0M in diethyl ether), ethylmagnesium bromide ($C_2H_5MgBr$, 1.0M in tetrahydrofuran), ethanethiol ($C_2H_5SH$), octanethiol ($C_8H_{17}SH$) and dodecanethiol ($C_{12}H_{25}SH$) were obtained from Sigma-Aldrich and used as received.

**Loading of Au seeds onto high surface area silica**. As-received silica particles (300mg) were first immersed in a 3-aminopropyltetraethylsilane (APTES) solution (12μL in 20mL methanol) for 2 min. After thorough rinsing with methanol to remove excess

APTES and dried in ambient air, the powder was soaked in pre-formed 20nm Au colloid solution (20mL, concentration $10^{12}$ particles per mL) for 10 min for electrostatic absorption of Au particles onto the APTES modified silica. The suspension was then centrifuged (at 3000 round-per-minute or rpm for 10 min) to remove any excess Au colloid particles in the supernatant. The silica supported Au seeds were then collected, dried and used for growth of Ge nanowires.

**Synthesis of Ge nanowires**. CVD growth of GeNWs on the silica supported Au seeds employed a condition similar to the growth of NWs on substrates reported recently. In a typical growth, 0.3g of silica powder loaded with 20nm growth seeds was placed in the center of a 1" quartz tube. The system was evacuated to its base pressure (~150mTorr) and heated up to growth temperature 295°C. $GeH_4$ (10% in He) was flowed into the system at a rate of 10 standard cubic centimeters per minute (sccm) and the pressure of the system was maintained at 50 Torr during the growth. After ~30 min of reactions, the $GeH_4$ feeding was stopped and the system was cooled in vacuum. The typical weight gain by the starting supported seed material after such growth is ~ 0.4g.

**Purification and separation of Ge nanowires from supported seed material.** After growth, the mixture of silica support and GeNWs was suspended in isopropanol (IPA) by a brief (10-20 s) sonication, during which GeNWs were detached from the support and became suspended in solution with the silica support particles precipitated out. The suspension was centrifuged (at 2000 rpm for 15 min) for collecting the suspended nanowires in the supernatant. The wires were further treated by soaking in a 5% HF solution for 10 min to remove any residual silica impurities, followed by DI $H_2O$ rinsing for 10 s and drying in a $N_2$ stream. It should be noted that no obvious nanowire corrosion was observed during the brief HF soaking process as TEM showed smooth nanowire surfaces after this treatment step. The length of the NWs after the 10-20 s sonication process was averaged at 8 μm (found by microscopy imaging). The time of the sonication was found to affect the wire length systematically. Longer sonication times tended to give shorter nanowires, although the final length distribution of the wires was highly dependent on the concentration of the wires and specific sonicators used.

**Functionalization of GeNWs by alkanethiol reactions.** The H-terminated[1] as-purified GeNWs after HF treatment were quickly transferred to a glove box ($N_2$ filled, $O_2$ level ≤ 1ppm) for functionalization reactions. Prior to usage, all containers were rinsed with hexane and baked in glove box at 220°C for 30 min to remove absorbed $H_2O$. This step is done for all of the functionalization reactions in the current work. 5mL of neat alkanethiol ($C_2$, $C_8$, and $C_{12}$) was added to immerse GeNWs and heated [2] to 80°C for 2 (for $C_2$ and $C_8$ thiols) or 24 h (for $C_{12}$ thiol - we found that 2h reaction for $C_{12}$ thiol failed to give comparable passivation effect as the 24 h reaction). Afterwards, the GeNWs were rinsed in the glove box with anhydrous tetrahydrofuran (THF) and then in ambient air with acetone, methanol and IPA to ensure removal of excess thiol.

**Functionalization of GeNWs by Grignard reactions.** For Grignard reactions, as-purified GeNWs were treated with 5% HCl solution for 5 min to terminate the surface with Cl.[2] After brief rinsing with DI $H_2O$ for ~ 10 s and drying, the wires were transferred to the glove box. 5mL of alkyl magnesium bromide ($C_2$, $C_8$ and $C_{12}$) was added to immerse GeNWs and the reaction was allowed to last from 6 h to 2 days at 80°C.[3] Afterwards, the GeNWs were rinsed with anhydrous THF in glove box, and then with tetrahydrofuran (THF), methanol and IPA in ambient air.

**Langmuir-Blodgett (LB) film assembly.** The functionalized GeNWs are highly hydrophobic due to alkyl terminations on the surfaces and are readily suspended in various organic solvents including chlorobenzene and chloroform at a concentration of ~10μg/mL. The suspension was used to spread GeNWs on subphase 5% (v/v) ethanol in $H_2O$ contained in a home-made 1×4" Teflon trough. When the organic suspension was added drop-wise to the subphase surface, GeNWs spread out as the organic solvent vaporized. After adding ~2mL of suspensions, we used a Teflon barrier to compress the floating GeNWs and condense them to form a close-packed LB monolayer of GeNWs in a ~1"×0.5" area. A Si or $SiO_2$/Si substrate was then dipped into the GeNW/subphase interface to transfer a GeNW film onto the substrate. We noticed that GeNWs in the film were aligned along the trough edge and Teflon barrier. This arrangement was maintained during film transfer.

**X-ray photoelectron spectroscopy.** XPS characterizations were carried out on a PHI Quantum 2000 scanning microprobe equipped with an angle-resolved hemispherical electron energy analyzer at a base system pressure of $1 \times 10^{-9}$ Torr. All measurements were performed using the Al $K_\alpha$ line with a photon-energy of 1486.6 eV. Charge compensation during acquisition was applied using the improved dual neutralization capability including an electron flood gun and an ion gun. The samples for XPS were mats of GeNWs deposited on Si substrates after various functionalization treatments.

**Transmission and scanning electron microscope.** The transmission electron microscope (TEM) used was a Philips CM20 with a working acceleration voltage of 200 kV. The scanning electron microscope (SEM) used was a FEI Siron field emission instrument operated at relatively low acceleration voltages of 1 kV to 5 kV to minimize charging effect.